\documentclass[pre,preprint,superscriptaddress]{revtex4}
\usepackage{amssymb,graphicx}

\begin{document}
\title{Universal statistics of the knockout tournament}
\author{Seung Ki Baek}
\email{seungki@pknu.ac.kr}
\affiliation{Department of Physics, Pukyong National University, 608-737 Busan,
Korea}
\author{Il Gu Yi}
\affiliation{BK21 Physics Research Division and Department of Physics,
Sungkyunkwan University, Suwon 440-746, Korea}
\author{Hye Jin Park}
\affiliation{BK21 Physics Research Division and Department of Physics,
Sungkyunkwan University, Suwon 440-746, Korea}
\author{Beom Jun Kim}
\email{beomjun@skku.edu}
\affiliation{BK21 Physics Research Division and Department of Physics,
Sungkyunkwan University, Suwon 440-746, Korea}

\begin{abstract}
We study statistics of the knockout tournament, where only the winner of a
fixture progresses to the next. We assign a real number called
competitiveness to each contestant and find that the resulting distribution
of prize money follows a power law with an exponent close to unity if the
competitiveness is a stable quantity and a decisive factor to win a match.
Otherwise, the distribution is found narrow. The existing observation of
power law distributions in various kinds of real sports tournaments
therefore suggests that the rules of those games are constructed in such a
way that it is possible to understand the games in terms of the contestants'
inherent characteristics of competitiveness.
\end{abstract}

\maketitle


Competition is a ubiquitous form of
social interaction for distributing limited
resources among a number of individuals, often regarded as the opposite of
cooperation. Competition has been a main tenet in economics where a
perfectly competitive equilibrium is proven Pareto-efficient as long as
there are no externalities and public goods.
Moreover, the notion of natural selection in biological evolution is often
understood as proving competition `natural'. For these reasons, although
competition results in growing tension across a society, most people
have taken it for granted as an organising principle of our society.

Recently, Deng et al.~\cite{deng} claimed universal power-law
distributions of scores and prize money by observing various kinds of
sports such as tennis, golf, football, badminton, and so on.
According to their extensive data analysis, the probability to find scores
or prize money greater than $k$ always decays as a power law
$P_>(k) \sim k^{-(\gamma-1)}$ with an exponential cutoff
where the power-law exponent $\gamma-1$ ranges
between 0.01 and 0.39 depending on sports. In addition,
they presented a knockout-tournament model to explain the observations.
This is an intriguing approach since the most organised forms of competition
are usually found in sports. It is also popular to
run a knockout tournament, consisting of successive rounds where
only a winner in each fixture progresses to the next round, because it is
an efficient procedure to find who is the best with a small number of
fixtures. In other words, Deng et al. hinted a direct connection between the
structure of competition and its consequences.
Physicists have already recognised sports as a fruitful research field:
Statistics of athletic records has been pioneered by Gembris et
al.~\cite{gembris} and Wergen et al.~\cite{krug}, for example,
and there have been attempts to even predict the limiting performances in
the long run~\cite{rad1}.
Sports ranking combinatorics has also been considered by Park and
Newman~\cite{jpark,jpark2}.
If we are to understand the dynamics governing high achievements in sports
careers, in particular,
one famous theory along this direction is called the Matthew ``rich get richer''
effect~\cite{matthew,detrend,prowess}: It says that a higher position
leads to a better chance to progress further in career, resulting in an
extremely skewed distribution. 
The spatial Poisson process to model this effect indeed explains such
behaviour with $\gamma \le 1$, which is found in some empirical data sets.
However, we should point out that many
factors of competition are hidden in the probability of progress, and that
the stochastic process is totally indifferent to individual characteristics as
written in Ecclesiastes: ``the race is not to the swift, but time and chance
happenth to them all.''

In this work, we instead focus on
statistical analysis of a specific system of competition, i.e., the knockout
tournament among inhomogeneous participants. Our main point is that a large
part of statistics is universal in the sense that it is independent of most
details of the game but already determined by the tournament structure.
Let us consider a player's number of wins
denoted by $n$, for example. When the tournament has been finished, the
distribution of $n$ denoted by $P(n)$ is always an exponentially decreasing
function of $n$. It is a purely geometric property of the tournament
tree independent of any details of the
game, loosely mapped to the critical percolation on a binary
tree~\cite{perc}. If the prize money is highly skewed towards the best
players, similarly to real sports tournaments, one can assume that
the prize money $k_n$ after winning $n$
rounds is also an exponential function of $n$, that is, $k_n \sim z^n$
(Fig.~\ref{fig:tree}).
Combining these two, one finds that the distribution $P(k) \sim k^{-\gamma}$
with
\begin{equation}
\gamma = (\log_2 z)^{-1}  + 1,
\label{eq:gamma}
\end{equation}
and this mechanism belongs to combination of exponentials according to
Newman~\cite{newman}.
If $z$ gets very large, $\gamma$ converges to unity, yielding $P(k) \sim
k^{-1}$. As $z \rightarrow 1$, on the other hand, $\gamma$ diverges because
$P(k)$ approaches the distribution function of $n$, which is an
exponential function.
In fact, if $z<2$, the total amount of prize money gets unbounded
as the number of contestants grows, which means that the organiser of this
tournament has a risk of bankruptcy. 
This explains why $k_n$ has to be such a rapidly increasing function of $n$,
and we see that the feasible range of $\gamma$ is between one and two.
Moreover, if there is a typical number of
prize winners, $z$ is effectively very large, driving $\gamma$ to
unity.
This is a simple prediction for a \emph{single} tournament. In other words,
this analysis corresponds to gathering data of prize money distributed
over many tournaments without identifying who was who. The actual
statistics
collected in this way, however, will not be very interesting to us, and it
is usually more meaningful to consider individual-based
statistics: Even for a team sport, each team may be regarded as
an individual. It is
notable that Deng et al. resolve this problem by introducing the
notion of ranks, belonging to individuals, and also by assuming that a
player's winning probability against another is a function of their rank
difference. Following this approach, we will see how our simple
prediction in equation (\ref{eq:gamma}) can be reproduced on average in the
individual-based statistics.


\section*{Results}
\subsection*{Decisiveness of competitiveness}
\label{sec:indi}

Imagine a tournament with $N=2^m$ contestants to construct a simple binary
tree.  Each person is assigned a real number $r$, which we refer 
as competitiveness instead of a rank, and reserve the
latter term for denoting an outcome of competition, which may or may not
reflect an individual's genuine competitiveness depending on how much luck
comes into play. By defining $r$ as a real number, the competitiveness
is automatically assumed to be transitive, which means that if contestant
$A$ is more competitive than $B$ who is more competitive than $C$, then $A$
is also more competitive than $C$. Since we can always rescale the highest
competitiveness as unity and the lowest one as null without loss of
generality, the real number $r$ belongs to a closed interval from zero to
one.

Under total uncertainty about the contestants,
we may assume as our initial condition that the distribution
of $r$ is uniformly random at the starting point. We thus denote the initial
probability density distribution of $r$ as $p_0(r) = 1$ with normalisation
$\int_0^1 p_0(r) dr = 1$.
Then, we introduce a function $f(r,r')$ that defines the probability for a
contestant with competitiveness $r$ to defeat another with
$r'$. As was done by Deng et al.~\cite{deng}, it can be assumed to be a
function of $x
\equiv r-r'$ only, and it is plausible in such a case that $f(x)$ is a
nondecreasing function of $x \in [-1,1]$ with $f(x)+f(-x)=1$.
In words, the former condition means that a more competitive player
has a higher probability to defeat a less competitive player, whereas the
latter condition is merely a simple reflection of the trivial fact that one
of the two players must win, irrespective of their values of $r$.
Let us check some examples of $f(x)$.

\subsubsection*{Perfect resolution}

One of the simplest choices is
\begin{equation}
f(r,r') = \Theta(r-r'),
\label{eq:step}
\end{equation}
where $\Theta$ is the Heaviside step function. This means that the
competitiveness decides the outcome deterministically.
In Methods, we have derived the following
nonlinear recursive relation
\begin{equation}
p_{n+1} (r) = 2 p_n(r) \int_0^1 dr' f(r,r') p_n(r'),
\label{eq:recur}
\end{equation}
where $p_n(r)$ means the distribution of $r$ after the $n$th round.
With the Heaviside step function, this equation is solvable at any arbitrary
$n$ and we obtain
\begin{equation}
p_n(r) = 2^n r^{2^n-1},
\label{eq:pn}
\end{equation}
with a corresponding cumulative distribution $c_n(r) \equiv \int_0^r
p_n(r') dr' = r^{2^n}$.
As explained in Methods, $c_n(r)$ is identical to
the winning chance for the contestant with $r$ at the $(n+1)$th round, denoted
by $w_n(r)$, when we have chosen the step function in equation (\ref{eq:step}).
 
We can extract various useful information from this probability density
function. For example, the average competitiveness after the $n$th round is
\begin{equation}
\left< r \right>_n = \int_0^1 dr~r p_n(r) = \frac{1}{1+2^{-n}},
\label{eq:avgr}
\end{equation}
and therefore the width of $p_n(r)$ decreases as $\sigma \sim 2^{-n}$.
A contestant with $r$ passes the $n$th round but not the next one with
probability
\begin{equation}
q_n(r) = \left[ \prod_{k=0}^{n-1} w_k(r) \right] [1-w_n(k)] =
r^{2^n-1} \left(1-r^{2^n} \right),
\label{eq:qn}
\end{equation}
where we have used $w_k = c_k$ and the sum over $n$ is normalised to
unity for
any $r$ between zero and one.
The average prize money for this person with $r$ can thus be
calculated as
\begin{equation}
\bar{k}(r) = \sum_{n=0}^\infty k_n q_n(r).
\end{equation}
As shown in Fig.~\ref{fig:qn}, $q_n$ has a peak at $n^\ast = \log_2 \left[-
\frac{1}{\log_2 r} \right]$ and the summations above can be approximated as
\begin{equation}
\bar{k}(r) \approx k_{n^\ast} q_{n^\ast} = \frac{k_{n^\ast}}{4r}.
\label{eq:kbar}
\end{equation}
If $k_n = z^n$, it means that $\bar{k}(r) \approx \frac{1}{4} z^{n^\ast}
\propto \left(-\frac{1}{\ln r} \right)^{\log_2 z}
\approx (1-r)^{-\log_2 z}$ in the vicinity of $r=1$. Note that we have
approximated $r$ as unity at the denominator of equation~(\ref{eq:kbar}).
Therefore, Zipf's
plot shows a power law with slope $-\log_2 z$, leading to 
$P(k) \sim  k^{-\gamma}$ with $\gamma = (\log_2 z)^{-1} + 1$ due to the relationship between Zipf's
plot and $P(k)$~\cite{park}.
This exactly coincides with equation
(\ref{eq:gamma}) derived for a single tournament. We have numerically
performed tournaments and the results confirm validity of our
analysis as shown in Fig.~\ref{fig:iia}, where the numerical calculations of
$c_5(r)$ and $\left< r \right>_n$ agree perfectly with the analytic results.
The detailed procedure of our simulation is explained in Methods.

\subsubsection*{Imperfect resolution}

As an opposite extreme case, let us consider a situation where
individual competitiveness is totally irrelevant to the outcome of a
match and only luck decides. In
other words, we assume a constant function $f(x) = 1/2$. If we start from
$p_0 = 1$, the winning chance here is $w_0(r) = \int_0^1 dr'~ f(r,r')
p_0(r') = 1/2$. Note that $w_0$ is not identical to the cumulative
distribution any more. The next round has a distribution $p_1(r) = 2w_0(r)
p_0(r) = 1$, and this pattern is repeated all the way leading to $p_n(r) =
1$ for every $n$. It is also straightforward to obtain the same result by
substituting the constant $f(x)=1/2$ into the recursive equation
(\ref{eq:recur}).
The resulting $P(k)$ is just the most likely distribution of
the prize money among the $N$ players, so the maximum entropy principle
tells us to maximise
\begin{equation}
H = - \sum_k P(k) \ln P(k) - \mu \sum_k k P(k),
\end{equation}
where the first term is Shannon entropy and $\mu$ represents a Lagrangian
multiplier for constraining the average prize money.
When $H$ is maximised, it does not change under variation in $P(k)$ to the
first order, and we thus have
\begin{equation}
0 = \delta H = - \delta P(k) \sum_k \left[ 1+\ln P(k) + \mu k \right],
\label{eq:var}
\end{equation}
which leads us to
$P(k) \sim \exp(-k/k_c)$ with a characteristic scale $k_c$.

This implies a tendency that $P(k)$ usually exhibits a power law with an
exponent close to unity but that randomness makes the tail shorter.
Suppose that $f(x)$ has a finite resolving power, quantified by a
characteristic width $\Gamma$
over which $f(x)$ rapidly increases. The Heaviside
step function corresponds to a limiting case of $\Gamma \rightarrow 0$.
We can predict the followings when $\Gamma$
is finite but sufficiently small: At
the beginning of the competition, the
width $\sigma$ of $p_n(r)$ is much greater than $\Gamma$,
so $f(x)$ effectively
serves as a step function. The above analysis shows that $\sigma$ decreases
as $2^{-n}$ so it becomes comparable with $\Gamma$ after $\nu \sim
\log_2 (1/\Gamma)$ rounds. Thereafter, the decrease of $\sigma$ slows down.
Finally, when $\sigma \ll \Gamma$ after many rounds, the survivors'
competitiveness is irrelevant and the outcomes are mostly determined by
pure luck. Therefore, a natural guess for $P(k)$ would be
\begin{equation}
P(k) \sim k^{-\gamma} \exp \left( -k/k_\Gamma \right),
\end{equation}
with $k_\Gamma \sim O(z^{\nu})$ and $\gamma$ in equation (\ref{eq:gamma}).
This functional form is confirmed in our numerical simulations
(Fig.~\ref{fig:iib}).
This distribution can also be derived from the maximum entropy principle as in
equation~(\ref{eq:var}) but with an additional constraint on $\sum_k \ln
k$~\cite{zipf,visser}, which
corresponds to the total number of fixtures in this context.
The above argument can be pursued further by employing the following $f(x)$:
\begin{equation}
f(x) = \left\{
\begin{array}{ll}
1-\frac{1}{2} e^{-x/\Gamma}  & \mbox{for ~} x>0, \\
\frac{1}{2} e^{x/\Gamma} & \mbox{otherwise,}
\end{array}\right.
\end{equation}
where the exponential functions make it possible to explicitly evaluate the
integral. Then, the winning chance is given as
\begin{eqnarray}
c_0(r) &=& \int_0^1 dr'~f(r,r')p_0(r')\\
&=& \int_0^r dr'~f(r,r')p_0(r') + \int_r^1 dr'~f(r,r')p_0(r')\\
&=& r + \frac{\Gamma}{2} e^{-r/\Gamma} - \frac{\Gamma}{2}
e^{(r-1)/\Gamma},
\end{eqnarray}
which approaches $c_0(r) = r$ as $\Gamma \rightarrow 0$ and $c_0(r) = 1/2$
as $\Gamma \rightarrow \infty$, as expected.
As above, this yields
\begin{equation}
p_1(r) = 2 c_0(r) p_0(r) = 2r + \Gamma e^{-r/\Gamma} - \Gamma
e^{(r-1)/\Gamma},
\label{eq:p1i}
\end{equation}
which is normalised to unity as $\int_0^1 dr~p_1(r) = 1$.
This result is quite suggestive, because
equation~(\ref{eq:p1i}) modifies equation (\ref{eq:pn}) at $n=1$ by adding
$O(\Gamma)$
when $r \lesssim \Gamma$ and subtracting the same amount when $(1-r)
\lesssim \Gamma$
[Fig.~\ref{fig:pn}(a)]. In
short, $p_0(r)$ becomes flatter when $r$ is close to $0$ or $1$.
If we take one step further, the low-$r$ correction becomes
less important and we find
\begin{eqnarray}
p_2(r) \approx 4r^3 - 6\Gamma r~ e^{(r-1)/\Gamma},
\end{eqnarray}
where we have left only the dominant correction of $O(\Gamma)$
[Fig.~\ref{fig:pn}(b)]. For general $n$,
the result up to the correction of $O(\Gamma)$ is inductively found as
\begin{eqnarray}
p_n(r) \approx 2^n r^{2^n-1} - 2^{n-1} \left(2^n-1 \right) \Gamma
r^{2^{n-1}-1} e^{(r-1)/\Gamma}.
\end{eqnarray}
This implies that the finite resolution is most
noticeable among highly competitive players with $(1-r) \lesssim
\Gamma$, whereas
the story looks similar to the case of perfect resolution when $(1-r)$ is
small but still much larger than $\Gamma$.

\subsection*{Stability of competitiveness}
\label{sec:update}

We have assumed that competitiveness is each individual's inherent
characteristic, which changes in a much longer time scale compared to
outcomes of
competition, and we relate the latter to ranks. The idea is that although
a contestant's
rank fluctuates over tournaments, it will correctly reflect her true
competitiveness in the long run. Even if the competitiveness may interact
with actual tournament results, it will usually be related to a cumulative
measure of performance
that mainly reflects low-frequency, i.e., long-term behaviour.
For example, we have calculated
the Kendall tau rank correlation coefficient~\cite{kendall}, denoted by $\tau$,
to see how the accumulated amounts of prize money change their
relative positions between two successive tournaments (Fig.~\ref{fig:tau}).
If a certain pair of contestants keep their relative positions,
they are said to be concordant, and discordant otherwise. The
coefficient $\tau$ is defined as the number of concordant pairs minus that of
discordant pairs, divided by the total number of possible pairs.
Beginning with the same initial amount of money for every contestant,
which is set to zero, we run fifty tournaments in a row, accumulating the
prize money for each individual. A contestant's accumulated money from
a series of tournaments determines her performance in the next
tournament in
such a way that $r=(N-i)/(N-1)$ is assigned to the contestant when she has
the $i$th largest accumulated amount. The relative positions of
two equal amounts are random.
In spite of this variability,
the ranks of the accumulated money get stabilised
after 20 or 30 tournaments in all the cases considered
(Fig.~\ref{fig:tau}), and the 
resulting $P(k)$ is almost identical to the static-$r$ case for each $\Gamma$.
Still, one may ask
what happens if their time scales approach each other so that a current rank
directly affects performance at the next tournament, provided that the
tournaments are regular events. Even if an individual's rank fluctuates
over time, it might still be possible for this correlation between successive
tournaments to reproduce the power-law tail part of $P(k)$.
In fact, this question
is not really well-posed because a knockout tournament leaves many contestants'
ranks undetermined except a few prize winners, and this is the fundamental
advantage of a knockout tournament. We nevertheless suppose
that a player's competitiveness at the next time step is a
nondecreasing function
of the current performance, say, $r_{t+1} = R(n_t)$, where $n_t$ is the
number of wins in the tournament at time $t$, and $R$ is a
nondecreasing function between zero and one. Since $r$
determines how many rounds the contestant can go through, the distribution
of $n_{t+1}$ is essentially a
function of $n_t$. The situation is actually boring
because the same contestant wins the first place all the time, but we may
exclude this exceptional contestant from our consideration.
We begin with noting that any tournament results in a distribution of
$n_t$ as $p_0(n_t) = 2^{-n_t-1}$, which is the initial
distribution of the next tournament at time $t+1$. The corresponding cumulative
portion of contestants with results below $n_t$ is thus
$c_0(n_t) = 1-2^{-n_t}$. As
above, if $f(x)$ is the Heaviside step function with $f(0) = 1/2$, the
chance to win the first round for a contestant that passed $n_t$ rounds at
the previous tournament is $w_0(n_t) = \frac{1}{2}p_0(n_t) + c_0(n_t)$. The
first term represents the probability to meet an opponent with the same
$n_t$, and the factor of one half originates from $f(0)$. The distribution
of $n_t$ at the next round is $p_1(n_t) = 2w_0(n_t) p_0(n_t)$. We can repeat
this procedure to obtain a general expression as
\begin{equation}
c_k (n_t) = \left[ 1- \frac{1}{g(n_t)} \right]^{g(k)}
\end{equation}
with $g(x) \equiv 2^x$. By definition, we have
\begin{equation}
p_k (n_t) = c_k (n_t +1) - c_k (n_t).
\end{equation}
If $k$ is not very small, $p_k (n_t)$
converges to a certain function of $y\equiv k-n_t$ with a maximum around
$y \approx 0$ [Fig.~\ref{fig:u}(a)]. The
conditional probability to reach $k$ and stop there for given $n_t$
is found as
\begin{equation}
q_k(n_t) = \left[ \prod_{j=0}^{k-1} w_j(n_t) \right] \left[
1-w_k(n_t) \right],
\end{equation}
with $\sum_{k=0}^\infty q_k(n_t) = 1$ [Fig.~\ref{fig:u}(b)].
We observe that $q_k(n_t)$ can
also be described as a certain function $V(y)$ when $n_t \gtrsim 3$.
Moreover, we find that $\sum_{k=0}^{n_t} q_k(n_t) > 1/2$ for any $n_t$.
In other words, the time series $\{n_t \ge 0\}$ can be roughly
described as a biased random walk towards the origin.
Since this holds true for anyone,
each contestant's average result
will be rapidly equalised by the bias so we predict that the probability
distribution $P(k)$ will be narrow. This prediction is well substantiated by
numerical results shown in Fig.~\ref{fig:iii}, where $P_>(k)$ is drawn in a
semi-log plot.
Therefore, in terms of the
time scale of competitiveness, the power-law shape of $P(k)$ is
observable when competitiveness changes much more slowly compared to
the frequency of tournaments.

\section*{Discussion}
\label{sec:conclusion}

In summary, we have investigated statistics resulting from knockout
tournaments.
It is basically the rules of the game that define competitiveness, so the
distribution of prize money is dependent on how much the rules take
individual competitiveness as a decisive and stable factor. But other
details of the game are found irrelevant, and the statistics is universal in
this sense.
More specifically, if competitiveness is a static parameter and any tiny
difference of it can
be distinguished by the rules, the distribution is predicted to
take a power-law shape $P(k) \sim k^{-\gamma}$ with $\gamma$ close to
unity.
If the difference is indistinguishable below a certain resolution limit
$\Gamma$,
we find an exponential cutoff at the tail, whose location is a function
of $\Gamma$.
We have also argued that the distribution $P(k)$ becomes narrow again when
competitiveness changes with a time scale comparable to the frequency of
tournaments. In this respect, the broad distributions observed across many
sports suggest that their rules are already stabilised in such a
way that one can readily compare contestants' competitiveness in a
consistent way over a long time span and that the result of competition
sensitively reflects the difference indeed. Since our analysis relates certain
internal parameters of a given tournament such as $z$ and $\Gamma$ to the final
distribution of prize money, which is somewhat more easily accessible, it will
an interesting question to verify such detailed relationships directly on
empirical grounds.
  
\section*{Methods}
\subsection*{Recursive relation for $p_n(r)$}
In case of perfect resolution, i.e., $f(r,r') = \Theta (r-r')$,
it is straightforward to obtain
the winning chance for the contestant with $r$ at the first round of
the tournament as
\begin{equation}
w_0(r) = \int_0^1 dr' f(r,r') p_0(r') = r,
\end{equation}
where $p_0(r') = 1$.
This happens to be identical to the cumulative distribution $c_0(r)$ and
it represents the simple fact
that the contestant with $r$ should meet an opponent with $r' < r$ in order
to win and progress to the next round.
When the first round has been finished,
the distribution of their competitiveness is
\begin{equation}
p_1(r) = 2 w_0(r) p_0(r) = 2r,
\label{eq:p1p}
\end{equation}
which is again normalised to unity.
The factor of two in front is needed because the number of survivors has
become one half of $N$.
Note that we have used independence between a
player's competitiveness and her opponent's in equation (\ref{eq:p1p}),
which is the case when the initial condition contains no correlations in
competitiveness.
As in the first round, the corresponding cumulative distribution,
\begin{equation}
c_1(r) = \int_0^1 dr'~f(r,r') p_1(r') = r^2,
\end{equation}
is identical to the winning chance $w_1(r)$ at the second round. In the same
way, the distribution after the second round is $p_2(r) = 2 w_1(r) p_1(r) =
4 w_1(r) w_0(r) p_0(r) = 4r^3$, and so on. For general $f(r,r')$, we can use
essentially the same argument to derive
the following nonlinear recursive relation:
\begin{equation}
p_{n+1} (r) = 2 p_n(r) \int_0^1 dr' f(r,r') p_n(r'),
\end{equation}
which is explicitly solvable for a few special cases as above.

\subsection*{Numerical procedures}
First, we generate a tournament tree with $N = 2^m$ contestants at the
terminal nodes and assign to each of them a real random number $r$ inside
the unit interval as competitiveness. One may require the minimum and
maximum of the random numbers to be strictly zero and one, respectively,
but it does not make a visible
difference when $N$ is large enough.
The resulting uncorrelated random number sequence
$\{ r_1, r_2, \ldots, r_N \}$ means
absence of a seeding process, so number one and number two seeds may face
each other in the first round. Second, when two contestants $A$ and $B$
meet with $r_A$
and $r_B$, respectively, we draw a random number $\rho \in [0,1)$ and choose
$A$ as the winner of this fixture if $\rho < f(r_A,r_B)$, and choose $B$
otherwise. This is repeated for every match
in this first round, and the winner progresses to the parent node. When we
have filled all the parent nodes with $2^{m-1}$ winners,
the second round starts among them in the same
way as before. As the tournament proceeds round by round, the number of
survivors decreases rapidly until the final winner is left alone after the
$m$th round. Each player defeated at the $n$th round receives prize money
$z^{n-1}$, whereas the final winner acquires $z^m$.
When a tournament is over, we start a new one with randomly shuffling $\{
r_1, r_2, \ldots, r_N \}$ at the terminal nodes, so that the competitiveness
is identified as an individual characteristic preserved across the
tournaments. We have performed $10^4$ shuffles, hence the same number of
tournaments, to obtain statistical averages for each $r_i$ with
$N=2^{12}$.


\begin{thebibliography}{10}
\expandafter\ifx\csname url\endcsname\relax
  \def\url#1{\texttt{#1}}\fi
\expandafter\ifx\csname urlprefix\endcsname\relax\def\urlprefix{URL }\fi
\providecommand{\bibinfo}[2]{#2}
\providecommand{\eprint}[2][]{\url{#2}}

\bibitem{deng}
\bibinfo{author}{Deng, W.}, \bibinfo{author}{Li, W.}, \bibinfo{author}{Cai,
  X.}, \bibinfo{author}{Bulou, A.} \& \bibinfo{author}{Wang, Q.~A.}
\newblock \bibinfo{title}{Universal scaling in sports ranking}.
\newblock \emph{\bibinfo{journal}{New J. Phys.}} \textbf{\bibinfo{volume}{14}},
  \bibinfo{pages}{093038} (\bibinfo{year}{2012}).

\bibitem{gembris}
\bibinfo{author}{Gembris, D.}, \bibinfo{author}{Taylor, J.~G.} \&
  \bibinfo{author}{Suter, D.}
\newblock \bibinfo{title}{Sports statistics: Trends and random fluctuations in
  athletics}.
\newblock \emph{\bibinfo{journal}{Nature (London)}}
  \textbf{\bibinfo{volume}{417}}, \bibinfo{pages}{506} (\bibinfo{year}{2002}).

\bibitem{krug}
\bibinfo{author}{Wergen, G.}, \bibinfo{author}{Volovik, D.},
  \bibinfo{author}{Redner, S.} \& \bibinfo{author}{Krug, J.}
\newblock \bibinfo{title}{Rounding effects in record statistics}.
\newblock \emph{\bibinfo{journal}{Phys. Rev. Lett.}}
  \textbf{\bibinfo{volume}{109}}, \bibinfo{pages}{164102}
  (\bibinfo{year}{2012}).

\bibitem{rad1}
\bibinfo{author}{Radicci, F.}
\newblock \bibinfo{title}{Universality, limits and predictability of gold-medal
  performances at the {O}lympic {G}ames}.
\newblock \emph{\bibinfo{journal}{PLoS ONE}} \textbf{\bibinfo{volume}{7}},
  \bibinfo{pages}{e40335} (\bibinfo{year}{2012}).

\bibitem{jpark}
\bibinfo{author}{Park, J.} \& \bibinfo{author}{Newman, M. E.~J.}
\newblock \bibinfo{title}{A network-based ranking system for {US} college
  football}.
\newblock \emph{\bibinfo{journal}{J. Stat. Mech.}}
  \textbf{\bibinfo{volume}{P10014}} (\bibinfo{year}{2005}).

\bibitem{jpark2}
\bibinfo{author}{Park, J.}
\newblock \bibinfo{title}{Diagrammatic perturbation methods in networks and
  sports ranking combinatorics}.
\newblock \emph{\bibinfo{journal}{J. Stat. Mech.}}
  \textbf{\bibinfo{volume}{P04006}} (\bibinfo{year}{2010}).

\bibitem{matthew}
\bibinfo{author}{Petersen, A.~M.}, \bibinfo{author}{Jung, W.-S.},
  \bibinfo{author}{Yang, J.-S.} \& \bibinfo{author}{Stanley, H.~E.}
\newblock \bibinfo{title}{Quantitative and empirical demonstration of the
  {M}atthew effect in a study of career longevity}.
\newblock \emph{\bibinfo{journal}{Proc. Natl. Acad. Sci. USA}}
  \textbf{\bibinfo{volume}{108}}, \bibinfo{pages}{18--23}
  (\bibinfo{year}{2011}).

\bibitem{detrend}
\bibinfo{author}{Petersen, A.~M.}, \bibinfo{author}{Penner, O.} \&
  \bibinfo{author}{Stanley, H.~E.}
\newblock \bibinfo{title}{Methods for detrending success metrics to account for
  inflationary and deflationary factors}.
\newblock \emph{\bibinfo{journal}{Eur. Phys. J. B}}
  \textbf{\bibinfo{volume}{79}}, \bibinfo{pages}{67--78}
  (\bibinfo{year}{2011}).

\bibitem{prowess}
\bibinfo{author}{Petersen, A.~M.}, \bibinfo{author}{Jung, W.-S.} \&
  \bibinfo{author}{Stanley, H.~E.}
\newblock \bibinfo{title}{On the distribution of career longevity and the
  evolution of home-run prowess in professional baseball}.
\newblock \emph{\bibinfo{journal}{EPL}} \textbf{\bibinfo{volume}{83}},
  \bibinfo{pages}{50010} (\bibinfo{year}{2008}).

\bibitem{perc}
\bibinfo{author}{Baek, S.~K.}, \bibinfo{author}{Minnhagen, P.} \&
  \bibinfo{author}{Kim, B.~J.}
\newblock \bibinfo{title}{Percolation on hyperbolic lattices}.
\newblock \emph{\bibinfo{journal}{Phys. Rev. E}} \textbf{\bibinfo{volume}{79}},
  \bibinfo{pages}{011124} (\bibinfo{year}{2009}).

\bibitem{newman}
\bibinfo{author}{Newman, M. E.~J.}
\newblock \bibinfo{title}{Power laws, {P}areto distributions and {Z}ipf's law}.
\newblock \emph{\bibinfo{journal}{Contemp. Phys.}}
  \textbf{\bibinfo{volume}{46}}, \bibinfo{pages}{323--351}
  (\bibinfo{year}{2005}).

\bibitem{park}
\bibinfo{author}{Kim, B.~J.} \& \bibinfo{author}{Park, S.~M.}
\newblock \bibinfo{title}{Distribution of {K}orean family names}.
\newblock \emph{\bibinfo{journal}{Physica A}} \textbf{\bibinfo{volume}{347}},
  \bibinfo{pages}{683--694} (\bibinfo{year}{2005}).

\bibitem{zipf}
\bibinfo{author}{Baek, S.~K.}, \bibinfo{author}{Bernhardsson, S.} \&
  \bibinfo{author}{Minnhagen, P.}
\newblock \bibinfo{title}{Zipf's law unzipped}.
\newblock \emph{\bibinfo{journal}{New J. Phys.}} \textbf{\bibinfo{volume}{13}},
  \bibinfo{pages}{043004} (\bibinfo{year}{2011}).

\bibitem{visser}
\bibinfo{author}{Visser, M.}
\newblock \bibinfo{title}{Zipf's law, power laws and maximum entropy}.
\newblock \emph{\bibinfo{journal}{New J. Phys.}} \textbf{\bibinfo{volume}{15}},
  \bibinfo{pages}{043021} (\bibinfo{year}{2013}).

\bibitem{kendall}
\bibinfo{author}{Kendall, M.}
\newblock \bibinfo{title}{A new measure of rank correlation}.
\newblock \emph{\bibinfo{journal}{Biometrika}} \textbf{\bibinfo{volume}{30}},
  \bibinfo{pages}{81--89} (\bibinfo{year}{1938}).

\end{thebibliography}

\acknowledgments
We are indebted to Petter Minnhagen for introducing us to this problem.
We thank Korea Institute for Advanced Study for providing computing
resources (KIAS Center for Advanced Computation, Abacus System) for this
work.
B.J.K. was supported by the National Research Foundation of Korea (NRF)
grant funded by the Korea government (MEST) (No. 2011-0015731).

\section*{Author contributions}
B.S.K. \& B.J.K. designed research, performed research, wrote, reviewed and
approved the manuscript. I.G.Y. \& H. J. P. performed the numerical and
statistical analysis of the data.

\section*{Additional information}
Competing financial interests: The authors declare no competing financial
interests.

\section*{Figure Legends}

\begin{figure}[!ht]
\caption{Schematic illustration of a tournament with four contestants $A$,
$B$, $C$, and $D$.
Contestant $B$ has competitiveness $r_B$ and gets prize money
$k_B = z^2$ because she has defeated $A$ and $C$. Likewise, $C$ gets $k_C =
z^1$ because she has won only a single match against $D$.}
\label{fig:tree}
\end{figure}

\begin{figure}[!ht]
\caption{Conditional probability to progress only to the
$n$th round for given competitiveness $r$ [see equation (\ref{eq:qn})].}
\label{fig:qn}
\end{figure}

\begin{figure}[!ht]
\caption{(a) Probability distribution of $r$ at the $5$th
round when $f(x)$ is the Heaviside step function, equation (\ref{eq:step}).
The data points are obtained numerically by
simulating $10^4$ tournaments with $N=2^{12}$ and the line shows our
analytic prediction in equation
(\ref{eq:pn}). (b) Average value of $r$ at the $n$th round, where the
data points are obtained numerically and the line represents
equation (\ref{eq:avgr}).}
\label{fig:iia}
\end{figure}

\begin{figure}[!ht]
\caption{Cumulative distribution of prize money, where
the horizontal axis is rescaled with respect to the largest value.
The data points are
obtained numerically by simulating $10^4$ tournaments with $N=2^{12}$
and $z=2$, in ascending order of $\Gamma$ from below.
The straight line shows our analytic prediction for $\Gamma=0$ for
comparison.}
\label{fig:iib}
\end{figure}

\begin{figure}[!ht]
\caption{Effects of imperfect resolution. (a) $p_1(r)$, the
distribution of competitiveness after the first round and (b) $p_2(r)$ after
the second round. The resolution parameter is the width of $f(x)$, which is
set to be
$\Gamma=5\%$ here. For comparison, the dotted lines show the cases
for $\Gamma=0$.}
\label{fig:pn}
\end{figure}

\begin{figure}[!ht]
\caption{Behaviour of the Kendall tau rank correlation
coefficient for the contestants' performance when the
each contestant's cumulative prize money determines her competitiveness.}
\label{fig:tau}
\end{figure}

\begin{figure}[!ht]
\caption{(a) The horizontal axis means the result of a tournament at time $t$,
and the vertical axis means probability to find a contestant with $n_t$ at
the $k$th round of the next tournament at $t+1$. Note the
similarity in shape at $k \gtrsim 4$, which means that $p_k(n_t) \approx
U(y)$ with $y\equiv k-n_t$. (b) Conditional probability $q_k(t)$ also
converges to a certain function $V(y)$ (see text).
}
\label{fig:u}
\end{figure}

\begin{figure}[!ht]
\caption{Cumulative distribution of prize money, when each
contestant's tournament result at time $t$ determines her competitiveness
at $t+1$. We have numerically simulating $10^4$ tournaments with $N=2^{12}$
and $z=2$. We have used the Heaviside step function as $f(x)$, and this plot
has excluded the one that always wins the first place.}
\label{fig:iii}
\end{figure}

\section{Figures}
\includegraphics[width=0.45\textwidth]{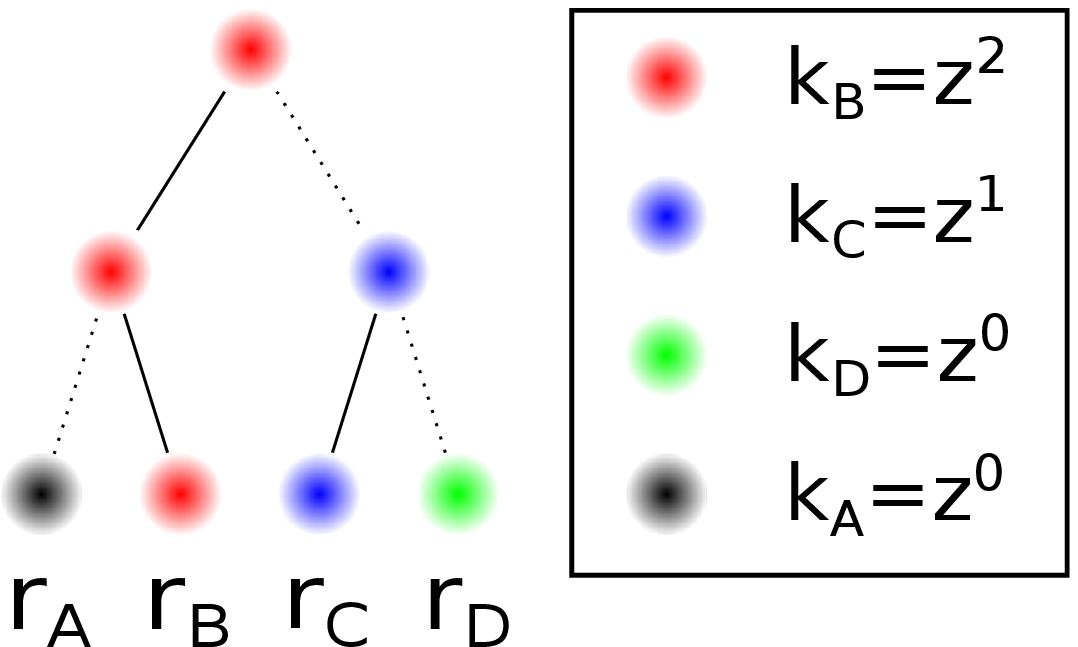}\\
\includegraphics[width=0.45\textwidth]{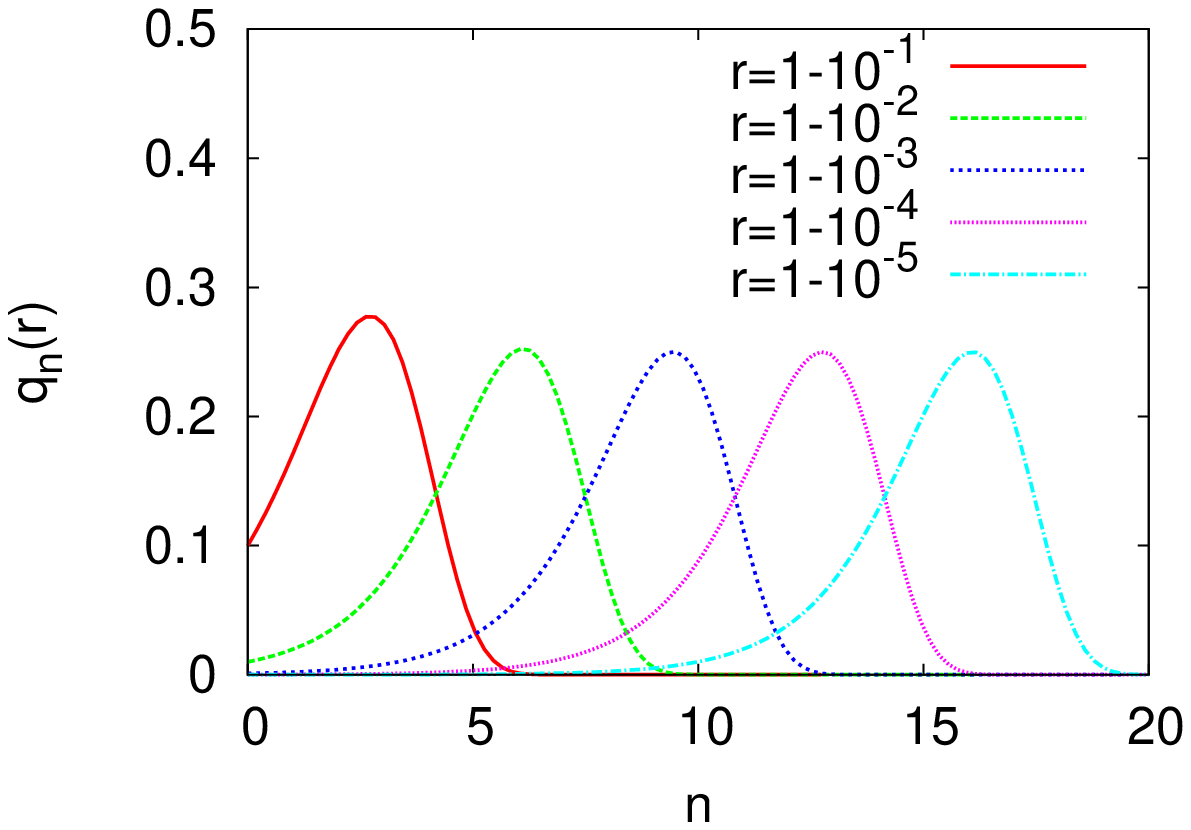}\\
\includegraphics[width=0.45\textwidth]{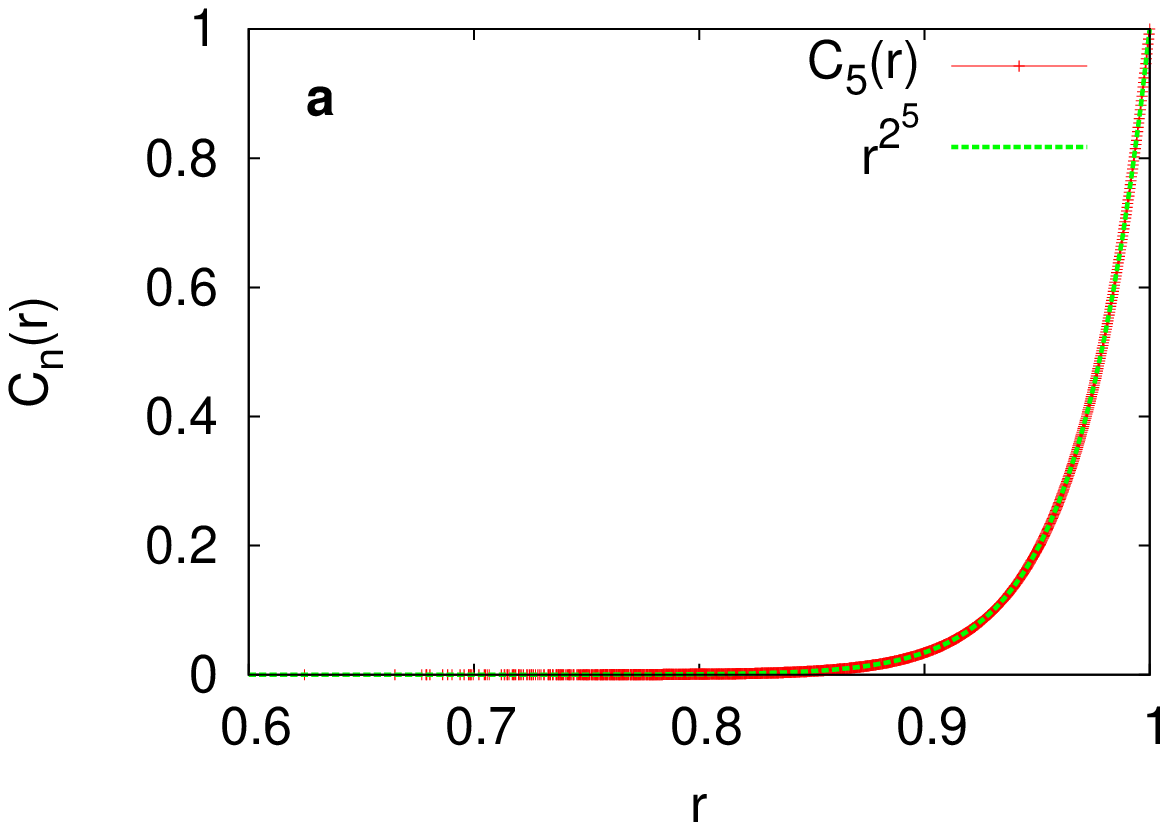}
\includegraphics[width=0.45\textwidth]{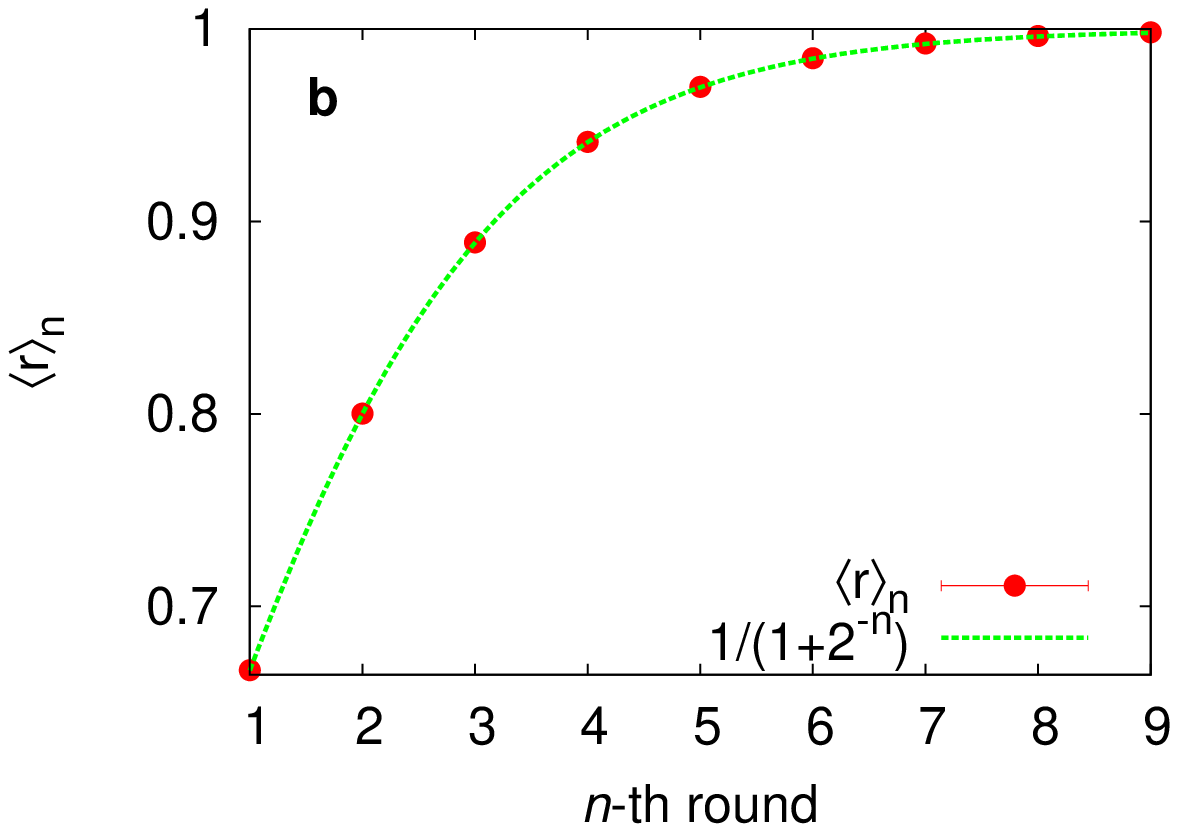}\\
\includegraphics[width=0.45\textwidth]{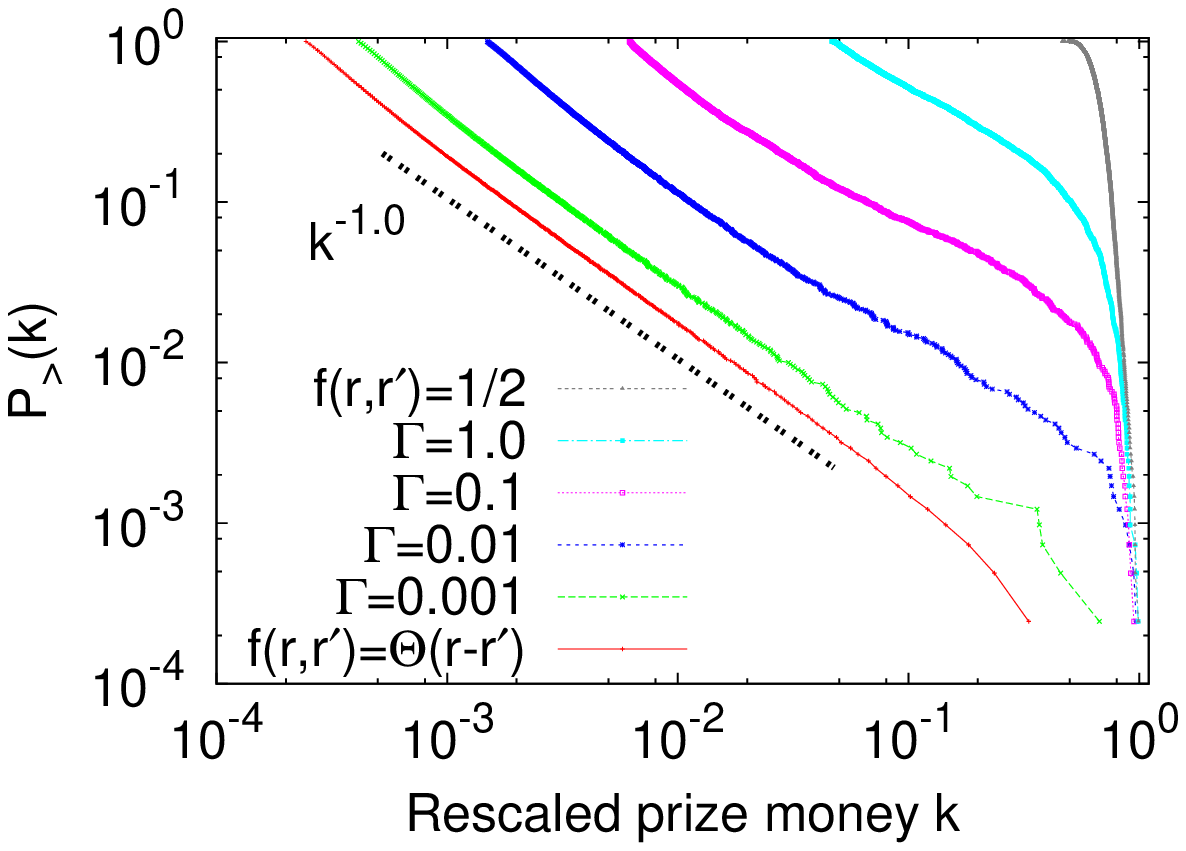}\\
\includegraphics[width=0.45\textwidth]{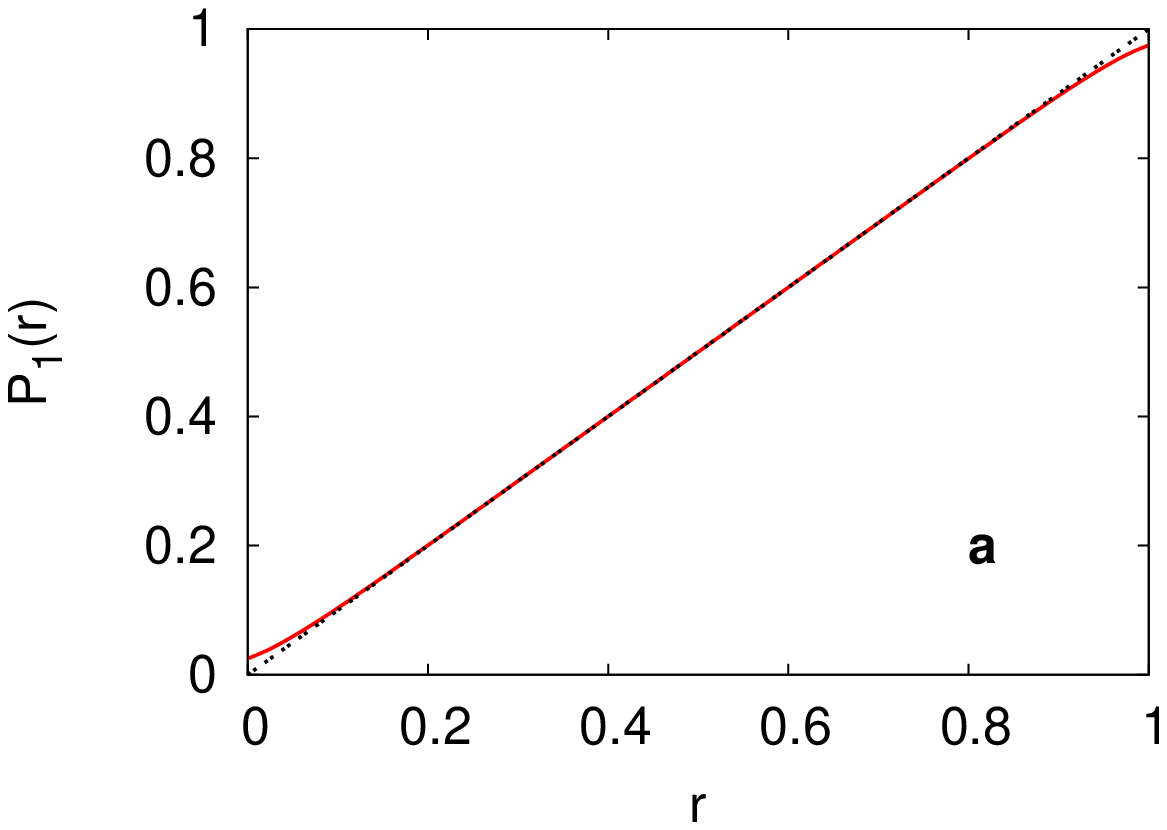}
\includegraphics[width=0.45\textwidth]{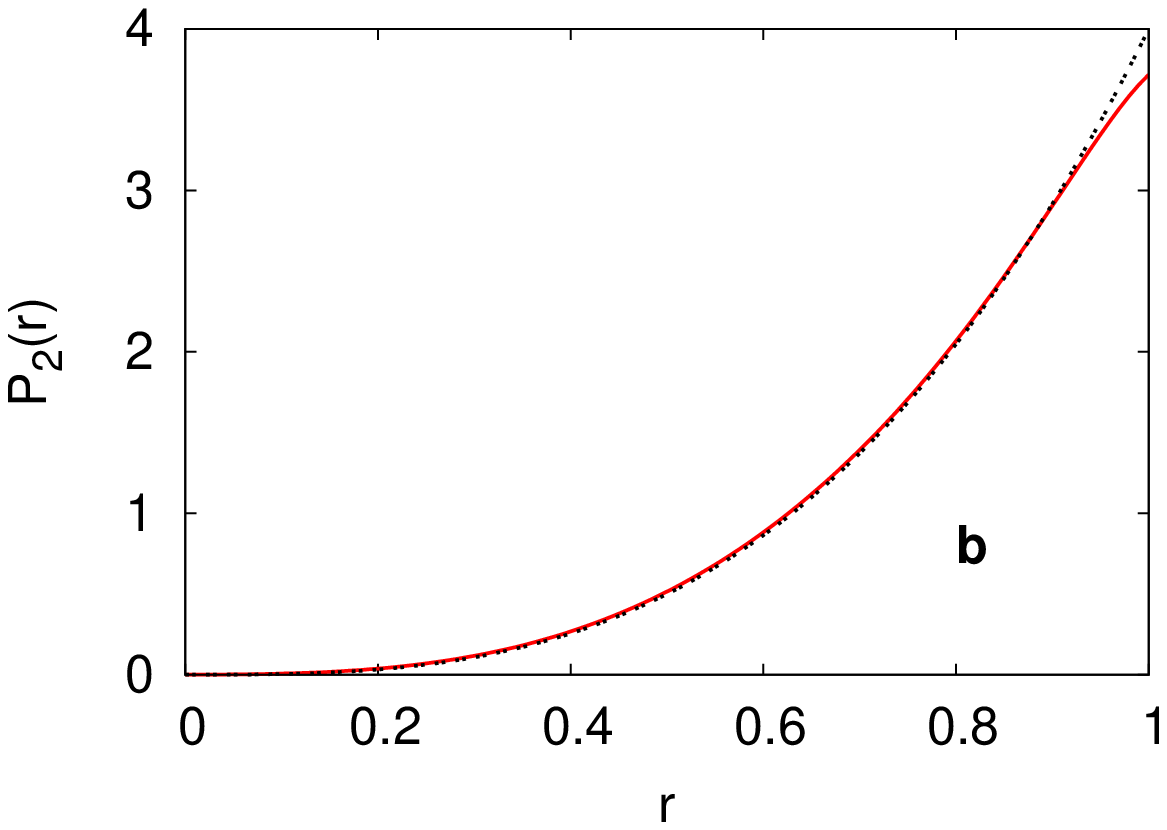}\\
\includegraphics[width=0.45\textwidth]{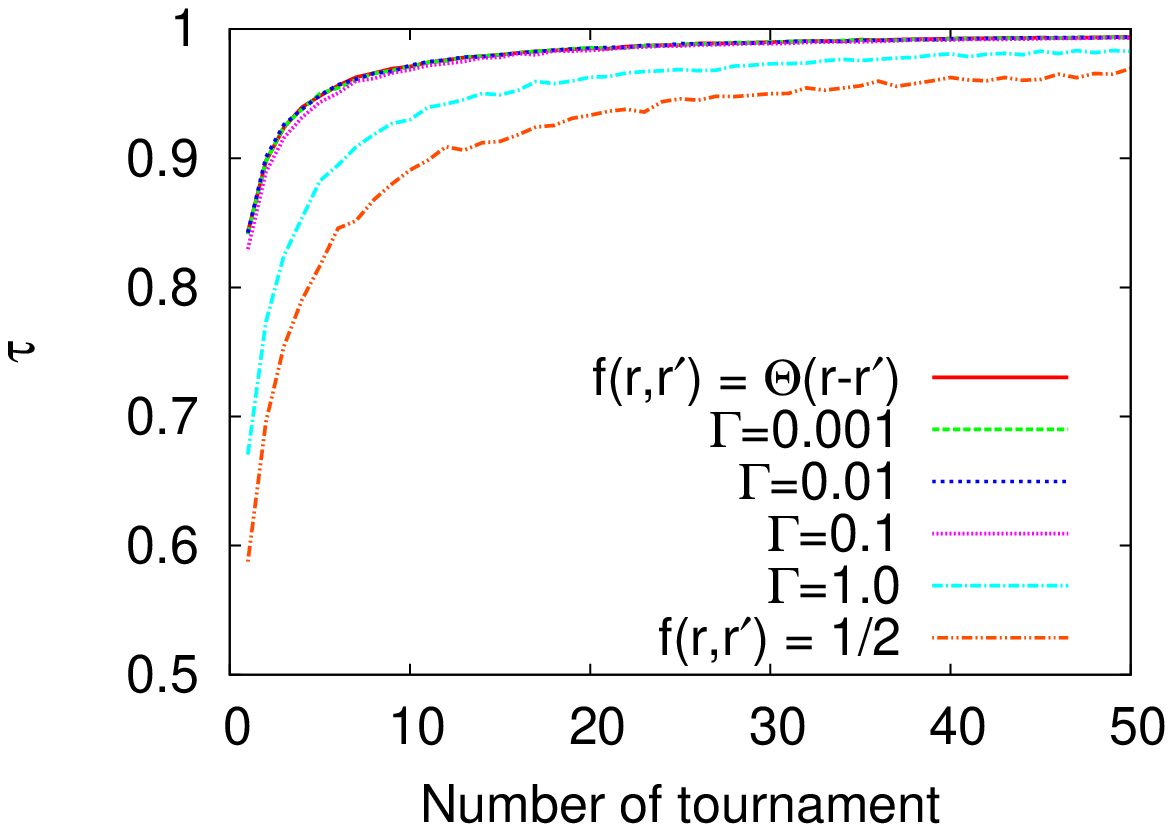}\\
\includegraphics[width=0.45\textwidth]{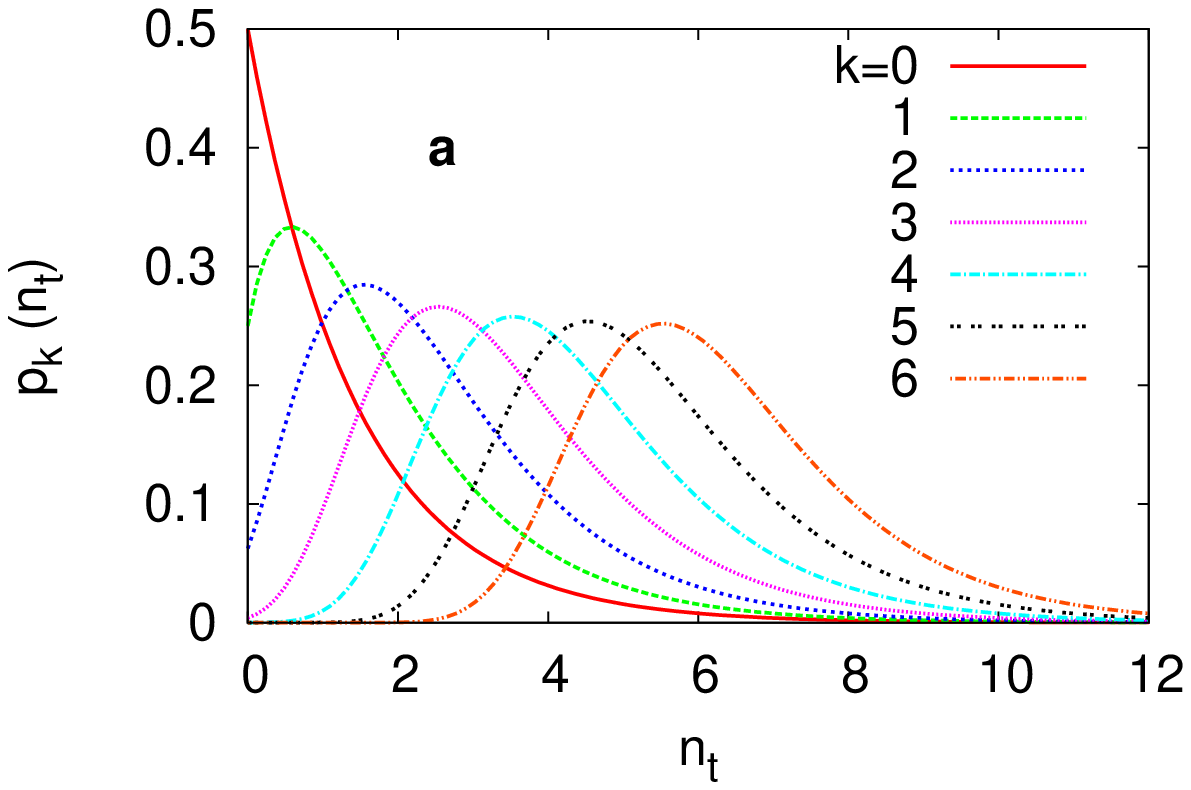}
\includegraphics[width=0.45\textwidth]{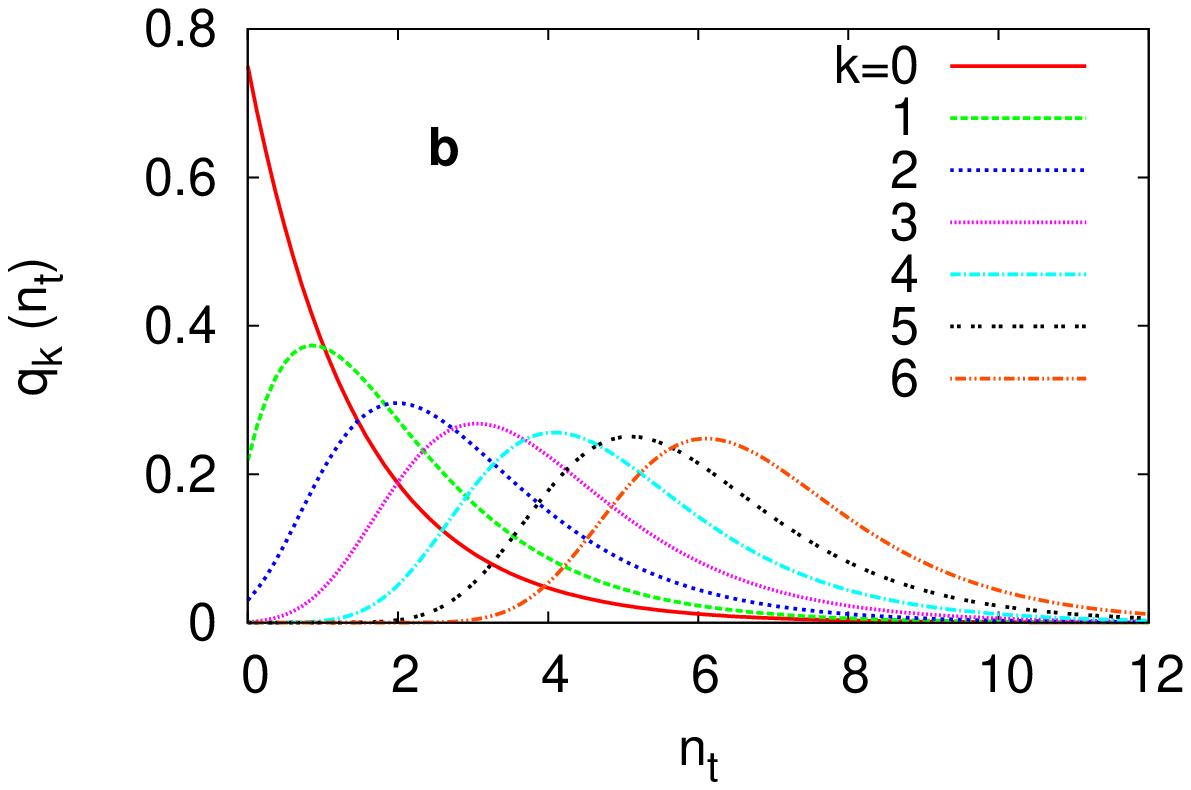}\\
\includegraphics[width=0.45\textwidth]{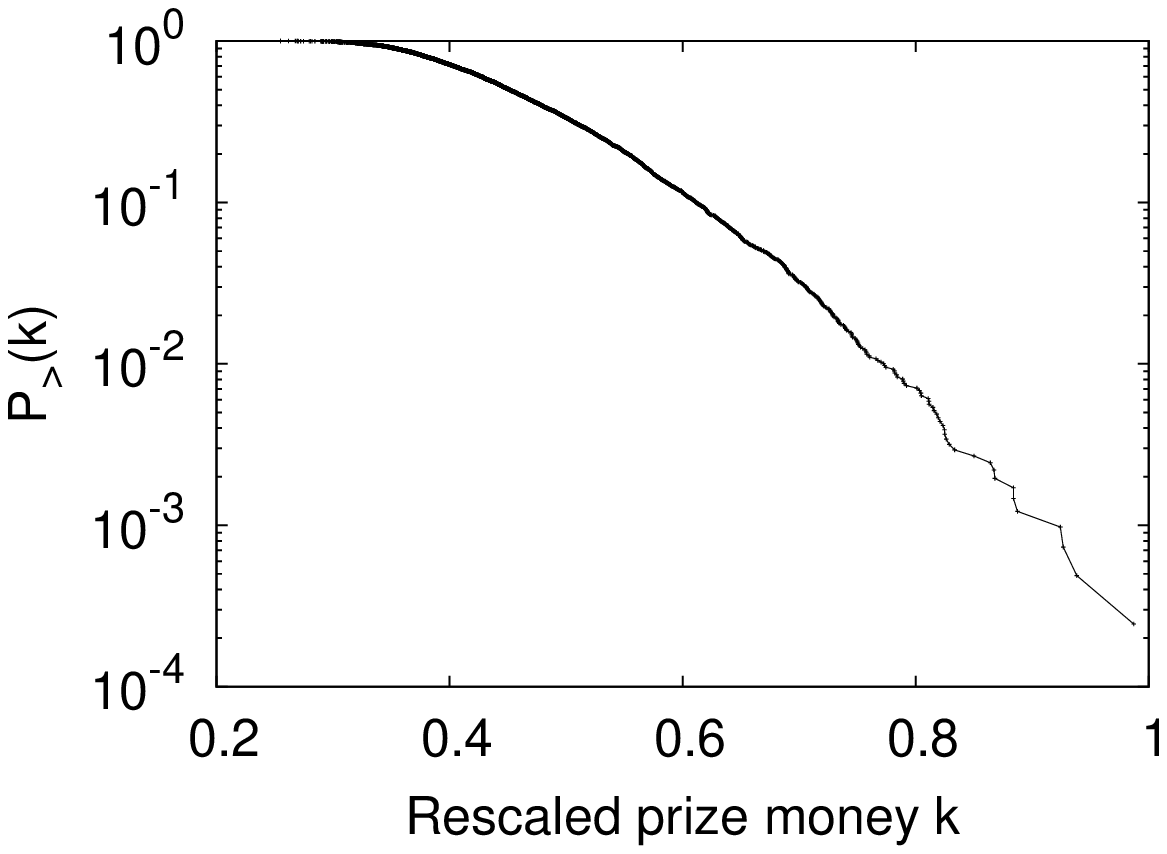}

\end{document}